\documentclass[12pt]{iopart}
\usepackage{iopams}  
\usepackage{graphicx} 
\begin{document}

\title[Density operators obtained through frame quantization]{Quantum states defined by using\\ the finite frame quantization}

\author{Nicolae Cotfas}
\address{Faculty of Physics, University of Bucharest, Romania}
\ead{ncotfas@yahoo.com , https://unibuc.ro/user/nicolae.cotfas/}
\vspace{10pt}
\begin{indented}
\item \today
\end{indented}

\begin{abstract}
Finite frame quantization is a discrete version of the coherent state quantization.
In the case of a quantum system with finite-dimensional Hilbert space, 
the finite frame quantization allows us to associate a linear operator 
to each function defined on the discrete phase space of the system.
We investigate the properties of the density operators which can be defined by using this method.
\end{abstract}

%
%
%
%
%

%
\section{Introduction}
The quantum particle moving along a straight line is described by using the Hilbert space $L^2(\mathbb{R})$.
For the corresponding classical system, $\mathbb{R}$ is the configuration space and $\mathbb{R}^2\!=\!\mathbb{R}\!\times\!\mathbb{R}$ the phase space. The  position operator
\begin{equation}
\hat q\psi (q)\!=\!q\, \psi(q)
\end{equation}
and the  momentum operator
\begin{equation}
\hat p\!=\!-{\rm i}\hbar \frac{d}{dq}
\end{equation}
satisfy the commutation relation
\begin{equation}
[\hat q,\hat p]\!=\!{\rm i}\hbar
\end{equation}
and the relation
\begin{equation}
\hat p \!=\!\hat F^\dag \hat q\hat F
\end{equation}
where
\begin{equation}
\hat F[\psi](p)\!=\!\mbox{\small $\frac{1}{\sqrt{2\pi\hbar}}$}\int\limits_{-\infty}^\infty {\rm e}^{-\frac{\rm i}{\hbar}pq}\psi(q)\, dq\!=\!\mbox{\small $\frac{1}{\sqrt{h}}$}\int\limits_{-\infty}^\infty {\rm e}^{-\frac{2\pi \rm i}{h}pq}\psi(q)\, dq
\end{equation}
is the  Fourier transform.

In the odd-dimensional case, $d\!=\!2s\!+\!1$, a discrete version can be obtained by using
\begin{equation}
\mathcal{R} \!=\!\{-s, -s\!+\!1,...,s\!-\!1,s\}
\end{equation}
as a configuration space, the Hilbert space (several representations are presented)
\begin{equation}
\mathcal{H}\!\equiv\!\mathbb{C}^d\!\equiv\!\ell^2(\mathcal{R})\!=\!\{ \, \psi\!:\!\mathcal{R}\rightarrow \mathbb{C}\,\}\!\equiv\!\left\{\psi\!:\!\mathbb{Z}\rightarrow \mathbb{C}\left| \mbox{\small $\begin{array}{r}
\psi(n\!+\!d)\!=\!\psi(n),\\[-1mm]
\mbox{for all } n\!\in\!\mathbb{Z}
\end{array}$}\!\!\!\right.\right\}
\end{equation}
with\\[-8mm]
\begin{equation}
\langle \psi,\varphi \rangle\!=\!\sum\limits_{n=-s}^s \overline{\psi(n)}\varphi(n)
\end{equation}
and the  {\it discrete Fourier} transform  $\hat{\frak{F}}\!:\!\mathcal{H}\!\rightarrow\!\mathcal{H}$,
\begin{equation}
\hat{\frak{F}}[\psi](k)\!=\!\mbox{\small $\frac{1}{\sqrt{d}}$} \sum\limits_{n=-s}^s{\rm e}^{-\frac{2\pi {\rm i}}{d}kn}\psi(n).
\end{equation}
The standard basis $\{ \delta_{-s},\ \delta_{-s+1},...,\delta_{s-1},\delta_s\}$, where
\begin{equation}
\delta_m(n)\!=\!\left\{ \begin{array}{lll}
1  & \mbox{if} & n\!=\!m\ \ \mbox{modulo } d\\[-1mm]
0  & \mbox{if} & n\!\neq\!m\ \ \mbox{modulo } d
\end{array} \right.
\end{equation}
is an orthonormal basis. By using Dirac's notation $|m\rangle$ instead of $\delta _m$, we have
\begin{equation}
\langle m|k \rangle \!=\!\delta_{mk},\qquad \sum\limits_{m=-s}^s|m\rangle\langle m|\!=\!\mathbb{I},
\end{equation}
where $\mathbb{I}\!:\!\mathcal{H}\!\rightarrow\!\mathcal{H}, \ \, \mathbb{I}\psi\!=\!\psi,$ \, is the identity operator.\\
In the discrete case, the  {\it position} operator $\hat{\frak{q}}:\mathcal{H}\!\rightarrow\!\mathcal{H}:\psi \!\mapsto\!\hat{\frak{q}}\psi$ is
\begin{equation}
\hat{\frak{q}}\psi(n)\!=\!n\, \psi(n).
\end{equation}
For the  {\it momentum} operator $\hat{\frak{p}}:\mathcal{H}\!\rightarrow\!\mathcal{H}$, the definition
\begin{equation}
\hat {\frak{p}} \!=\!\hat{\frak{F}}^\dag \hat{\frak{q}}\hat{\frak{F}}
\end{equation}
is more adequate than the use of a finite-difference operator instead of $\frac{d}{dq}$.\\
In the discrete case, the set
\begin{equation}
\mathcal{R}^2\!=\!\mathcal{R}\!\times\!\mathcal{R}\!=\!\{\, (n,k)\ |\ n,k\!\in\!\{-s, -s\!+\!1,...,s\!-\!1,s\}\, \}
\end{equation}
plays the role of phase space.

The Gaussian function of continuous variable ($\kappa \!>\!0$ is a parameter)
\begin{equation}
g_\kappa\!:\!\mathbb{R}\!\rightarrow\!\mathbb{R},\qquad g_\kappa(q)\!=\!{\rm e}^{-\frac{\kappa}{2\hbar}q^2}\!=\!{\rm e}^{-\frac{\kappa\pi }{h}q^2}
\end{equation}
satisfies the relation
\begin{equation}
\hat{F}[g_\kappa ]\!=\!\mbox{\small $\frac{1}{\sqrt{\kappa}}$}g_{\frac{1}{\kappa}}.
\end{equation}
The corresponding  {\it Gaussian function} of discrete variable, defined as \cite{Me,Ru}
\begin{equation}
\frak{g}_\kappa\!:\!\mathcal{R}\!\rightarrow\!\mathbb{R},\qquad 
\frak{g}_\kappa(n)\!=\!\sum\limits_{\alpha=-\infty}^\infty{\rm e}^{-\frac{\kappa \pi}{d}(n+\alpha d)^2}
\end{equation}
satisfies the similar relation 
\begin{equation}
\hat{\frak{F}}[\frak{g}_\kappa ]\!=\!\mbox{\small $\frac{1}{\sqrt{\kappa}}$}\frak{g}_{\frac{1}{\kappa}}.
\end{equation}
In this article, we restrict us to the odd-dimensional case, but most of the definitions and results can be extended in order to include the even-dimensional case $d\!=\!2s$ also.
\section{Coherent state quantization}
In the continuous case, the quantum state
\begin{equation}
|0,0\rangle\!=\!\mbox{\small $\frac{1}{\sqrt{\langle g_1,g_1\rangle}}$}|g_1\rangle 
\end{equation}
represents the {\it vacuum state}. The {\it coherent states} \cite{Pe}
\begin{equation}\label{ccs}
|q,p\rangle\!=\!\hat D(q,p)|0,0\rangle ,
\end{equation}
defined by using the {\it displacement operators} \cite{Pe,Sc}
\begin{equation}
\hat D(q,p)\!=\!{\rm e}^{-\frac{\rm i}{2\hbar}pq}\,{\rm e}^{\frac{\rm i}{\hbar}p\hat q}\,
{\rm e}^{-\frac{\rm i}{\hbar}q\hat p}\!=\!{\rm e}^{-\frac{\pi \rm i}{h}pq}\,{\rm e}^{\frac{2\pi {\rm i}}{h}p\hat q}\,
{\rm e}^{-\frac{2\pi {\rm i}}{h}q\hat p}
\end{equation}
satisfy the resolution of the identity
\begin{equation}
\mathbb{I}\!=\!\mbox{\small $\frac{1}{2\pi\hbar}$}\int\limits_{\mathbb{R}^2} |q,p\rangle\langle q,p|\, dqdp.
\end{equation}
By using the coherent state quantization,  we associate the linear operator \cite{Ga}
\begin{equation}
\hat A_f\!=\!\mbox{\small $\frac{1}{2\pi\hbar}$}\int\limits_{\mathbb{R}^2}f(q,p)\, |q,p\rangle\langle q,p|\, dqdp
\end{equation}
to each function  
\begin{equation}
f\!:\!\mathbb{R}\!\times\mathbb{R}\!\rightarrow \!\mathbb{C},
\end{equation}
defined on the phase space $\mathbb{R}^2$, and such that the integral is convergent.\\
For example, in the case $f(q,p)\!=\!q$, we get \cite{Ga}
\begin{equation}
\hat A_f\!=\!\mbox{\small $\frac{1}{2\pi\hbar}$}\int\limits_{\mathbb{R}^2}q\, |q,p\rangle\langle q,p|\, dqdp\!=\!\hat q,
\end{equation}
in the case $f(q,p)\!=\!p$, we get \cite{Ga}
\begin{equation}
\hat A_f\!=\!\mbox{\small $\frac{1}{2\pi\hbar}$}\int\limits_{\mathbb{R}^2}p\, |q,p\rangle\langle q,p|\, dqdp\!=\!\hat p,
\end{equation}
and, in the case $f(q,p)\!=\!\frac{p^2+q^2}{2}$, we get \cite{Ga}
\begin{equation}
\hat A_f\!=\!\mbox{\small $\frac{1}{2\pi\hbar}$}\int\limits_{\mathbb{R}^2}\frac{p^2\!+\!q^2}{2}\, |q,p\rangle\langle q,p|\, dqdp\!=\!-\frac{\hbar^2}{2}\frac{{\rm d}^2}{{\rm d}q^2}\!+\!\frac{1}{2}q^2\!+\!\frac{1}{2}.
\end{equation}
In the last case, the operator 
\begin{equation}
\hat A_f\!-\!\frac{1}{2}\!=\!-\frac{\hbar^2}{2}\frac{{\rm d}^2}{{\rm d}q^2}\!+\!\frac{1}{2}q^2
\end{equation}
is the Hamiltonian of the quantum harmonic oscillator.
\section{Finite frame quantization}
The quantum state \cite{Me,Ru}
\begin{equation}
|0;\!0\rangle\!=\!\mbox{\small $\frac{1}{\sqrt{\langle \frak{g}_1,\frak{g}_1\rangle}}$}|\frak{g}_1\rangle 
\end{equation}
can be regarded as a discrete counterpart of the {\it vacuum state} and $\hat{\frak{D}}(n,k)\!:\!\mathcal{H}\!\rightarrow\!\mathcal{H},$
\begin{equation}
\hat{\frak{D}}(n,k)\!=\!{\rm e}^{-\frac{\pi {\rm i}}{d}nk}\,{\rm e}^{\frac{2\pi {\rm i}}{d}k\hat \frak{q}}\,
{\rm e}^{-\frac{2\pi {\rm i}}{d}n\hat \frak{p}}
\end{equation}
as  {\it displacement operators} \cite{Sc,Vo}.\\[5mm]
{\bf Theorem 1.} {\it The discrete coherent states \cite{CGV,Vo}
\begin{equation}\label{dcs}
\qquad \qquad \qquad |n;\!k\rangle\!=\!\hat{\frak{D}}(n,k)|0;\!0\rangle ,
\end{equation}
\qquad \qquad \qquad satisfy the resolution of the identity
\begin{equation}\label{ResId}
\qquad \qquad \qquad \mathbb{I}\!=\!\mbox{\small $\frac{1}{d}$}\sum\limits_{n,k=-s}^s |n;\!k\rangle\langle n;\!k|.
\end{equation} }
\noindent{\it Proof.} Since
\[
\begin{array}{rl}
{\rm e}^{-\frac{2\pi {\rm i}}{d}n\hat \frak{p}}\frak{g}_1(m) & \!\!\!=\!{\rm e}^{-\frac{2\pi {\rm i}}{d}n\hat{\frak{F}}^\dag \hat{\frak{q}}\hat{\frak{F}}}\frak{g}_1(m)\!=\!\hat{\frak{F}}^\dag {\rm e}^{-\frac{2\pi {\rm i}}{d}n\hat{\frak{q}}}\hat{\frak{F}}\frak{g}_1(m)\\
 & \!\!\!=\!\frac{1}{\sqrt{d}}\sum\limits_{a=-s}^s{\rm e}^{\frac{2\pi {\rm i}}{d}ma} {\rm e}^{-\frac{2\pi {\rm i}}{d}n\hat{\frak{q}}}\hat{\frak{F}}\frak{g}_1(a)\\
 & \!\!\!=\!\frac{1}{\sqrt{d}}\sum\limits_{a=-s}^s{\rm e}^{\frac{2\pi {\rm i}}{d}ma} {\rm e}^{-\frac{2\pi {\rm i}}{d}na}\hat{\frak{F}}\frak{g}_1(a)\\
 & \!\!\!=\!\frac{1}{\sqrt{d}}\sum\limits_{a=-s}^s{\rm e}^{\frac{2\pi {\rm i}}{d}ma} {\rm e}^{-\frac{2\pi {\rm i}}{d}na}\frac{1}{\sqrt{d}}\sum\limits_{b=-s}^s{\rm e}^{-\frac{2\pi {\rm i}}{d}ab} \frak{g}_1(b)\\
 & \!\!\!=\!\sum\limits_{b=-s}^s\frac{1}{d}\sum\limits_{a=-s}^s{\rm e}^{\frac{2\pi {\rm i}}{d}a(m-n-b)} \frak{g}_1(b)\\
 & \!\!\!=\!\sum\limits_{b=-s}^s\delta_b(m\!-\!n) \frak{g}_1(b)\!=\!\frak{g}_1(m\!-\!n)\\
\end{array}
\]
and
\[
\begin{array}{rl}
\langle m|n;\!k\rangle & \!\!\!=\!\langle m|\hat{\frak{D}}(n,k)|0;\!0\rangle\\
 & \!\!\!=\!\mbox{\small $\frac{1}{\sqrt{\langle \frak{g}_1,\frak{g}_1\rangle}}$}{\rm e}^{-\frac{\pi {\rm i}}{d}nk}\,{\rm e}^{\frac{2\pi {\rm i}}{d}k\hat \frak{q}}\,
{\rm e}^{-\frac{2\pi {\rm i}}{d}n\hat \frak{p}}\frak{g}_1(m)\\
 & \!\!\!=\!\mbox{\small $\frac{1}{\sqrt{\langle \frak{g}_1,\frak{g}_1\rangle}}$}{\rm e}^{-\frac{\pi {\rm i}}{d}nk}\,{\rm e}^{\frac{2\pi {\rm i}}{d}km}\,
{\rm e}^{-\frac{2\pi {\rm i}}{d}n\hat \frak{p}}\frak{g}_1(m)\\
 & \!\!\!=\!\mbox{\small $\frac{1}{\sqrt{\langle \frak{g}_1,\frak{g}_1\rangle}}$}{\rm e}^{-\frac{\pi {\rm i}}{d}nk}\,{\rm e}^{\frac{2\pi {\rm i}}{d}km}\, \frak{g}_1(m\!-\!n)\\
\end{array}
\]
we get
\[
\begin{array}{rl}\fl
\left\langle m\left|\mbox{\small $\frac{1}{d}$}\sum\limits_{n,k=-s}^s |n;\!k\rangle\langle n;\!k| \right|\ell\right\rangle & \!\!\!=\!\frac{1}{d}\frac{1}{\langle \frak{g}_1,\frak{g}_1\rangle}
\sum\limits_{n,k=-s}^s{\rm e}^{\frac{2\pi {\rm i}}{d}km}\, \frak{g}_1(m\!-\!n)
{\rm e}^{-\frac{2\pi {\rm i}}{d}k\ell}\, \frak{g}_1(\ell\!-\!n)\\
 & \!\!\!=\!\frac{1}{\langle \frak{g}_1,\frak{g}_1\rangle}\sum\limits_{n=-s}^s\mbox{\small $\frac{1}{d}$}
\sum\limits_{k=-s}^s{\rm e}^{\frac{2\pi {\rm i}}{d}k(m-\ell)}\, \frak{g}_1(m\!-\!n)\, \frak{g}_1(\ell\!-\!n)\\
 & \!\!\!=\!\frac{1}{\langle \frak{g}_1,\frak{g}_1\rangle}\sum\limits_{n=-s}^s\delta_{m\ell}\, \frak{g}_1(m\!-\!n)\, \frak{g}_1(\ell\!-\!n)\!=\!\delta_{m\ell}.\qquad \Box\\

\end{array}
\]

\newpage

By using the finite frame quantization, we associate the linear operator \cite{CG,CGV,Ga}
\begin{equation}\label{FrameQuant}
\hat{\Lambda}_f\!=\!\mbox{\small $\frac{1}{d}$}\sum\limits_{n,k=-s}^s f(n,k)\, |n;\!k\rangle\langle n;\!k|
\end{equation}
to each function  
\begin{equation}
f\!:\!\mathcal{R}\!\times\!\mathcal{R}\!\rightarrow \!\mathbb{C}
\end{equation}
defined on the discrete phase space $\mathcal{R}^2\!=\!\mathcal{R}\!\times\!\mathcal{R}$.\\[3mm]
{\bf Theorem 2.} {\it Let $f,g\!:\!\mathcal{R}\!\times\!\mathcal{R}\!\rightarrow \mathbb{C}$ and $\alpha,\beta\!\in\!\mathbb{C}$. We have:}
\begin{equation}
a) \quad f(n,k)\!=\!1\quad \Rightarrow \quad \hat{\Lambda}_f\!=\!\mathbb{I}.
\end{equation}
\begin{equation}
b) \quad\hat{\Lambda}_{\alpha f+\beta g}\!=\!\alpha \hat{\Lambda}_f\!+\!\beta \hat{\Lambda}_g.
\end{equation}
\begin{equation}
c) \quad \begin{array}{c}
 f(n,k)\!\in\!\mathbb{R}\\[-1mm]
\mbox{\it for any }n,k
\end{array}\quad \Rightarrow \quad \hat{\Lambda}_f^\dag\!=\!\hat{\Lambda}_f.
\end{equation}
\begin{equation}
d) \quad \begin{array}{c}
 f(n,k)\!\geq\!0\\[-1mm]
\mbox{\it for any }n,k
\end{array}\quad \Rightarrow \quad \hat{\Lambda}_f\!\geq\!0.
\end{equation}
\begin{equation}
e) \quad {\rm tr}\,\hat{\Lambda}_f\!=\!\mbox{\small $\frac{1}{d}$}\sum\limits_{n,k=-s}^s f(n,k).
\end{equation}
{\it Proof.}\\
a) \ Direct consequence of (\ref{ResId}).\\
b) \ Direct consequence of the definition (\ref{FrameQuant}).\\[-7mm]
\[\fl
\begin{array}{rl}
\!\!\!c)\quad\hat{\Lambda}_f^\dag\!= & \!\!\!\mbox{\small $\frac{1}{d}$}\sum\limits_{n,k=-s}^s \overline{f(n,k)}\, (|n;\!k\rangle\langle n;\!k|)^\dag\\
= & \!\!\!\mbox{\small $\frac{1}{d}$}\sum\limits_{n,k=-s}^s f(n,k)\, |n;\!k\rangle\langle n;\!k|\!=\!\hat{\Lambda}_f.
\end{array}
\]
d)\quad For any $\psi \!\in\!\mathcal{H}$, we have
\[\fl
\quad \ \ \begin{array}{rl}
\langle \psi , \hat{\Lambda}_f\psi\rangle\!= & \!\!\!\mbox{\small $\frac{1}{d}$}\sum\limits_{n,k=-s}^s f(n,k)\, \langle \psi|n;\!k\rangle\langle n;\!k|\psi\rangle\\
=  & \!\!\!\mbox{\small $\frac{1}{d}$}\sum\limits_{n,k=-s}^s f(n,k)\, |\langle n;\!k|\psi\rangle|^2\!\geq\!0.\\
\end{array}
\]
\[\fl
\begin{array}{rl}
\!\!\!e)\quad  {\rm tr}\,\hat{\Lambda}_f\!= & \!\!\!\sum\limits_{m=-s}^s \langle m|\hat{\Lambda}_f|m\rangle\\
= & \!\!\!\mbox{\small $\frac{1}{d}$}\sum\limits_{m=-s}^s \sum\limits_{n,k=-s}^s f(n,k)\,  \langle m|n;\!k\rangle\langle n;\!k|m\rangle\\
= & \!\!\!\mbox{\small $\frac{1}{d}$}\sum\limits_{n,k=-s}^s f(n,k)\sum\limits_{m=-s}^s  \langle n;\!k|m\rangle \langle m|n;\!k\rangle\\
= & \!\!\!\mbox{\small $\frac{1}{d}$}\sum\limits_{n,k=-s}^s f(n,k) \langle n;\!k|\mathbb{I}|n;\!k\rangle\\
= & \!\!\!\mbox{\small $\frac{1}{d}$}\sum\limits_{n,k=-s}^s f(n,k) .\qquad \Box
\end{array}
\]

In the case $f(n,k)\!=\!\frac{n^2\!+\!k^2}{2}$, the operator $ \hat{\Lambda}_f\!-\!\frac{1}{2}$ can be regarded as a discrete version of the Hamiltonian of the quantum harmonic oscillator. The eigenfunctions $\psi_n$ of $ \hat{\Lambda}_f$, considered in the increasing order of the number of sign alternations, can be regarded as a finite counterpart of the Hermite-Gauss functions $\Psi_n(q)$. In the cases analyzed in \cite{CD2}, the eigenfunctions $\psi_n$ of $ \hat{\Lambda}_f$ approximate $\Psi_n(q)$ better than the Harper functions $\frak{h}_n$, and approximately satisfy the relation \\[-7mm]
\begin{equation}
\hat \frak{F}[\psi_n]\!=\!(-{\rm i})^n\psi_n.\\[-5mm]
\end{equation}
Instead of the standard definition of the {\it discrete fractional Fourier transform} \cite{Ba,Ca,OZK}\\[-7mm]
\begin{equation}
\hat \frak{F}^\alpha\!=\!\sum\limits_{n=0}^{d-1}(-{\rm i})^{n\alpha}|\frak{h}_n\rangle\langle \frak{h}_n|.
\end{equation}
one can use \cite{CD2}\\[-7mm]
\begin{equation}
\hat \frak{F}^\alpha\!=\!\sum\limits_{n=0}^{d-1}(-{\rm i})^{n\alpha}|\psi_n\rangle\langle \psi_n|.
\end{equation}
as an alternative definition. The Harper functions (available only numerically) are defined as the eigenfunctions of a discrete version of the Hamiltonian of the quantum harmonic oscillator obtained by using finite-differences \cite{Ba,Ca,OZK}. The finite frame quantization \cite{CD2,CG,CGV} seems to behave better than the method based on finite-differences when we have to obtain discrete versions of certain operators.
\section{Density operators obtained through finite frame quantization}
The finite frame quantization allows us to define a remarkable class of quantum states.\\
{\bf Theorem 3.} {\it If the function $f\!:\!\mathcal{R}\!\times\!\mathcal{R}\!\rightarrow [0,d]$ is such that}\\[-7mm]
\begin{equation}
\qquad \sum\limits_{n,k=-s}^s f(n,k)\!=\!d,
\end{equation}
\qquad \qquad \qquad {\it then the corresponding linear operator $\hat \varrho_f\!:\!\mathcal{H}\!\rightarrow\!\mathcal{H}$,}\\[-7mm]
\begin{equation}\label{fdensity}
\qquad \hat \varrho_f\!=\!\mbox{\small $\frac{1}{d}$}\sum\limits_{n,k=-s}^s f(n,k)\, |n;\!k\rangle\langle n;\!k|
\end{equation}
\qquad \qquad \qquad {\it is a density operator}.\\
{\it Proof.} Direct consequence of theorem 2.$\qquad \Box$\\[5mm]
For example, the state corresponding to $f(n,k)\!=\!\frac{1}{d}$ is the mixed state $\hat \varrho_f\!=\!\frac{1}{d}\mathbb{I}$,\\ and the state corresponding to 
\begin{equation}\label{fcoherent}
f(n,k)\!=\!\left\{\begin{array}{lll}
d & \mbox{for} & (n,k)\!=\!(m,\ell)\\
0 & \mbox{for} & (n,k)\!\neq\!(m,\ell)
\end{array}\right.
\end{equation}
is the pure state $\hat \varrho_f\!=\!|m;\!\ell\rangle\langle m;\!\ell|$, that is, the discrete coherent state $|m;\!\ell\rangle$.\\[3mm]
{\bf Theorem 4.} {\it The set $\mathcal{S}_{\rm fr}$ of all the density operators of the form (\ref{fdensity}) is a convex set.}\\
{\it Proof.} 
If $\lambda\!\in\![0,1]$ and $\hat \varrho_f,\hat \varrho_g\!\in\!\mathcal{S}_{\rm fr}$, then
\[
(1\!-\!\lambda)\hat \varrho_f\!+\!\lambda \hat \varrho_g\!=\!\hat \varrho_h,\quad \mbox{where}\quad 
h(n,k)\!=\!(1\!-\!\lambda)f(n,k)\!+\!\lambda g(n,k). \qquad \Box
\]
{\bf Theorem 5.} {\it $\mathcal{S}_{\rm fr}$ is the convex hull of the set of pure states $\{\, |n;\!k\rangle\langle n;\!k|\ |\ n,k\!\in\!\mathcal{R}\,\}$.}\\
{\it Proof.} The purity of a state $\hat \varrho_f$ is\\[-7mm]
\begin{equation}
{\rm tr}\,\hat \varrho_f^2\!=\!\mbox{\small $\frac{1}{d^2}$}\sum\limits_{n,k=-s}^s\sum\limits_{m,\ell=-s}^s f(n,k)\,f(m,\ell)\, |\langle n;\!k|m;\!\ell\rangle |^2.
\end{equation}
\mbox{}\\[-6mm]
Since\\[-8mm]
\begin{equation}
 |\langle n;\!k|m;\!\ell\rangle |^2\!\leq\! \langle n;\!k|n;\!k\rangle\, \langle m;\!\ell|m;\!\ell\rangle\!=\!1,
\end{equation}
$\hat \varrho_f$ is a pure state if and only if $f$ is a function of the form (\ref{fcoherent}), that is, $\hat \varrho_f$ is one of the discrete coherent states $|m;\!\ell\rangle\langle m;\!\ell|.\qquad \Box$\\[3mm]
{\bf Theorem 6.} {\it If the function $f\!:\!\mathcal{R}\!\times\!\mathcal{R}\!\rightarrow [0,d]$ is such that}\\[-7mm]
\begin{equation}
\qquad \sum\limits_{n,k=-s}^s f(n,k)\!=\!d,
\end{equation}
\qquad \qquad \qquad {\it then the mean value}
\begin{equation}
\qquad \left\langle \hat A\right\rangle_{\!\!\hat \varrho_f}\!=\! {\rm tr}(A\, \hat \varrho_f)
\end{equation}
\qquad \qquad \qquad {\it of an observable $\hat A\!:\!\mathcal{H}\!\rightarrow\!\mathcal{H}$ in the state $\hat \varrho_f$ is}\\[-7mm]
\begin{equation}
\qquad \left\langle \hat A\right\rangle_{\!\!\hat \varrho_f}\!=\!\mbox{\small $\frac{1}{d}$}\sum\limits_{n,k=-s}^s f(n,k)\,\langle n;\!k|\hat A |n;\!k\rangle .
\end{equation}
\mbox{}\\[-4mm]
{\it Proof.} We have\\[-8mm]
\[
\qquad \begin{array}{rl}
\!\!\! \left\langle \hat A\right\rangle_{\!\!\hat \varrho_f}\!= & \!\!\!\mbox{\small $\frac{1}{d}$}\sum\limits_{n,k=-s}^s f(n,k) \ {\rm tr}(\hat A\, |n;\!k\rangle\langle n;\!k|)\\
= & \!\!\!\mbox{\small $\frac{1}{d}$}\sum\limits_{n,k=-s}^s f(n,k) \sum\limits_{m=-s}^s \langle m|\hat A|n;\!k\rangle\langle n;\!k|m\rangle\\
= & \!\!\!\mbox{\small $\frac{1}{d}$}\sum\limits_{n,k=-s}^s f(n,k) \sum\limits_{m=-s}^s \langle n;\!k|m\rangle\langle m|\hat A|n;\!k\rangle\\
= & \!\!\!\mbox{\small $\frac{1}{d}$}\sum\limits_{n,k=-s}^s f(n,k) \, \langle n;\!k|\hat A|n;\!k\rangle .\qquad \Box
\end{array}
\]
{\bf Theorem 7.} {\it If the function $f\!:\!\mathcal{R}\!\times\!\mathcal{R}\!\rightarrow [0,d]$ is such that}\\[-7mm]
\begin{equation}
\qquad\qquad  \sum\limits_{n,k=-s}^s f(n,k)\!=\!d,
\end{equation}
\mbox{}\\[-4mm]
\qquad \qquad \qquad {\it then, under the Fourier transform, $\hat \varrho_f$ maps as}\\[-8mm]
\begin{equation}
\qquad\qquad \hat \varrho_f\ \mapsto \ \hat{\frak{F}}\hat \varrho_f\hat{\frak{F}}^\dag\!=\!\hat \varrho_g,
\end{equation}
\qquad \qquad \qquad {\it where} \ $g(n,k)\!=\!f(-k,n)$.\\
{\it Proof.} Since\\[-7mm]
\[
\begin{array}{rl}
\langle m|\hat{\frak{F}}|n;\!k\rangle\!= & \!\!\!\langle m|\hat{\frak{F}}\hat{\frak{D}}(n,k)|0;\!0\rangle\!=\!
\mbox{\small $\frac{1}{\sqrt{\langle \frak{g}_1,\frak{g}_1\rangle}}$}\,\hat{\frak{F}}\hat{\frak{D}}(n,k)\frak{g}_1(m) \\
= & \!\!\!\mbox{\small $\frac{1}{\sqrt{\langle \frak{g}_1,\frak{g}_1\rangle}}$}\,\mbox{\small $\frac{1}{\sqrt{d}}$} \sum\limits_{n=-s}^s{\rm e}^{-\frac{2\pi {\rm i}}{d}m\ell}\hat{\frak{D}}(n,k)\frak{g}_1(\ell)\\
= & \!\!\!\mbox{\small $\frac{1}{\sqrt{\langle \frak{g}_1,\frak{g}_1\rangle}}$}\,\mbox{\small $\frac{1}{\sqrt{d}}$} \sum\limits_{n=-s}^s{\rm e}^{-\frac{2\pi {\rm i}}{d}m\ell}{\rm e}^{-\frac{\pi {\rm i}}{d}nk}\,{\rm e}^{\frac{2\pi {\rm i}}{d}k\ell}\, \frak{g}_1(\ell\!-\!n)\\
= & \!\!\!\mbox{\small $\frac{1}{\sqrt{\langle \frak{g}_1,\frak{g}_1\rangle}}$}\,{\rm e}^{-\frac{\pi {\rm i}}{d}nk}\,\mbox{\small $\frac{1}{\sqrt{d}}$} \sum\limits_{n=-s}^s{\rm e}^{-\frac{2\pi {\rm i}}{d}\ell(m-k)}\, \frak{g}_1(\ell\!-\!n)\\
= & \!\!\!\mbox{\small $\frac{1}{\sqrt{\langle \frak{g}_1,\frak{g}_1\rangle}}$}\,{\rm e}^{-\frac{\pi {\rm i}}{d}nk}\,\mbox{\small $\frac{1}{\sqrt{d}}$} \sum\limits_{n=-s}^s{\rm e}^{-\frac{2\pi {\rm i}}{d}(\ell+n)(m-k)}\, \frak{g}_1(\ell)\\
= & \!\!\!\mbox{\small $\frac{1}{\sqrt{\langle \frak{g}_1,\frak{g}_1\rangle}}$}\,{\rm e}^{-\frac{\pi {\rm i}}{d}nk}\,{\rm e}^{-\frac{2\pi {\rm i}}{d}n(m-k)}\,\mbox{\small $\frac{1}{\sqrt{d}}$} \sum\limits_{n=-s}^s{\rm e}^{-\frac{2\pi {\rm i}}{d}\ell(m-k)}\, \frak{g}_1(\ell)\\
= & \!\!\!\mbox{\small $\frac{1}{\sqrt{\langle \frak{g}_1,\frak{g}_1\rangle}}$}\,{\rm e}^{\frac{\pi {\rm i}}{d}nk}\,{\rm e}^{-\frac{2\pi {\rm i}}{d}nm}\, \hat{\frak{F}}[\frak{g}_1](m\!-\!k)\\
= & \!\!\!\mbox{\small $\frac{1}{\sqrt{\langle \frak{g}_1,\frak{g}_1\rangle}}$}\,{\rm e}^{\frac{\pi {\rm i}}{d}nk}\,{\rm e}^{-\frac{2\pi {\rm i}}{d}nm}\, \frak{g}_1(m\!-\!k)\!=\!\langle m|k;\!-n\rangle,\\
\end{array}
\]
we have $ \hat{\frak{F}}|n;\!k\rangle\!=\!|k;\!-n\rangle$, and consequently\\[-7mm]
\[
\begin{array}{rl}
\hat{\frak{F}}\hat \varrho_f\hat{\frak{F}}^\dag\!= & \!\!\!\mbox{\small $\frac{1}{d}$}\sum\limits_{n,k=-s}^s f(n,k)\, \hat{\frak{F}}|n;\!k\rangle\langle n;\!k|\hat{\frak{F}}^\dag\\
= & \!\!\!\mbox{\small $\frac{1}{d}$}\sum\limits_{n,k=-s}^s f(n,k)\, |k;\!-\!n\rangle\langle k;\!-\!n|\\
= & \!\!\!\mbox{\small $\frac{1}{d}$}\sum\limits_{n,k=-s}^s f(-k,n)\, |n;\!k\rangle\langle n;\!k|.\qquad \Box
\end{array}
\]
{\bf Theorem 8.} {\it If the function $f\!:\!\mathcal{R}\!\times\!\mathcal{R}\!\rightarrow [0,d]$ is such that}\\[-7mm]
\begin{equation}
\qquad\qquad  \sum\limits_{n,k=-s}^s f(n,k)\!=\!d,
\end{equation}
\mbox{}\\[-4mm]
\mbox{}\qquad \qquad \qquad  {\it then, under the displacement $ \hat{\frak{D}}(m,\ell)$, the operator $\hat \varrho_f$ maps as}\\[-6mm]
\begin{equation}
\qquad\qquad \hat \varrho_f\ \mapsto \  \hat{\frak{D}}(m,\ell)\hat \varrho_f\hat{\frak{D}}^\dag(m,\ell)\!=\!\hat \varrho_g,
\end{equation}
\mbox{}\qquad \qquad \qquad {\it where} \ $g(n,k)\!=\!f(n\!-\!m\, ({\rm mod}\, d),k\!-\!\ell\, ({\rm mod}\, d))$.\\
{\it Proof.} We have (see \cite{C2})\\[-7mm]
\[\fl
\begin{array}{rl}
\hat{\frak{D}}(m,\ell)\hat \varrho_f\hat{\frak{D}}^\dag(m,\ell)\!= & \!\!\!\mbox{\small $\frac{1}{d}$}\sum\limits_{n,k=-s}^s f(n,k)\, \hat{\frak{D}}(m,\ell)|n;\!k\rangle\langle n;\!k|\hat{\frak{D}}^\dag(m,\ell)\\
& \!\!\!\!\!\!\!\!\!\!\!\!\!\!\!\!\!\!\!\!\!\!\!\!\!\!\!\!\!\!\!\!\!\!\!\!=\mbox{\small $\frac{1}{d}$}\sum\limits_{n,k=-s}^s f(n,k)\, |n\!+\!m\, ({\rm mod}\, d);\!k\!+\!\ell\, ({\rm mod}\, d)\rangle\langle n\!+\!m\, ({\rm mod}\, d);\!k\!+\!\ell\, ({\rm mod}\, d)|\\
& \!\!\!\!\!\!\!\!\!\!\!\!\!\!\!\!\!\!\!\!\!\!\!\!\!\!\!\!\!\!\!\!\!\!\!\!=\mbox{\small $\frac{1}{d}$}\sum\limits_{n,k=-s}^s f(n\!-\!m\, ({\rm mod}\, d),k\!-\!\ell\, ({\rm mod}\, d))\, |n;\!k\rangle\langle n;\!k|.\qquad \Box
\end{array}
\]
{\bf Theorem 9.} {\it If the function $f\!:\!\mathcal{R}\!\times\!\mathcal{R}\!\rightarrow [0,d]$ is such that}\\[-7mm]
\begin{equation}
\qquad\qquad  \sum\limits_{n,k=-s}^s f(n,k)\!=\!d,
\end{equation}
\mbox{}\\[-4mm]
\mbox{}\qquad \quad \ {\it then, under the transposition map $|j\rangle\langle \ell|\mapsto |\ell \rangle\langle j|$, the operator $\hat \varrho_f$ transforms as}\\[-7mm]
\begin{equation}
\qquad\qquad \hat \varrho_f\ \mapsto \  \hat \varrho_f^{T}\!=\!\hat \varrho_g,
\end{equation}
\mbox{}\\[-7mm]
\mbox{}\qquad  \qquad {\it where} \ $g(n,k)\!=\! f(n,-k)$.\\
{\it Proof.} Since\\[-7mm]
\[
|n;k\rangle \!=\!\sum\limits_{j=-s}^s|j\rangle\langle j|n;k\rangle \!=\!\sum\limits_{j=-s}^s|j\rangle\,
\mbox{\small $\frac{1}{\sqrt{\langle \frak{g}_1,\frak{g}_1\rangle}}$}\, {\rm e}^{-\frac{\pi {\rm i}}{d}nk}\,{\rm e}^{\frac{2\pi {\rm i}}{d}kj}\, \frak{g}_1(j\!-\!n),
\]
under the transposition map $|j\rangle\langle \ell|\mapsto |\ell \rangle\langle j|$, the operator
\[
|n;k\rangle \langle n;k| \!=\!\sum\limits_{j,\ell=-s}^s|j\rangle\langle \ell|\,
\mbox{\small $\frac{1}{\langle \frak{g}_1,\frak{g}_1\rangle}$}\, {\rm e}^{\frac{2\pi {\rm i}}{d}kj}\,{\rm e}^{-\frac{2\pi {\rm i}}{d}k\ell}\, \frak{g}_1(j\!-\!n)\, \frak{g}_1(\ell\!-\!n)
\] 
transforms to
\[
\sum\limits_{j,\ell=-s}^s|\ell \rangle\langle j|
\mbox{\small $\frac{1}{\langle \frak{g}_1,\frak{g}_1\rangle}$}{\rm e}^{\frac{2\pi {\rm i}}{d}kj}\,{\rm e}^{\frac{2\pi {\rm i}}{d}k\ell}\, \frak{g}_1(j\!-\!n)\, \frak{g}_1(\ell\!-\!n)\!=\!|n;-k\rangle \langle n;-k|.\qquad \Box
\] 
{\bf Theorem 10.} {\it If the function $f\!:\!\mathcal{R}\!\times\!\mathcal{R}\!\rightarrow [0,d]$ is such that}
\begin{equation}
\qquad\qquad  \sum\limits_{n,k=-s}^s f(n,k)\!=\!d,
\end{equation}
\mbox{}\\[-4mm]
\mbox{}\qquad \qquad  {\it then, under the parity transform $|j\rangle\mapsto \Pi|j\rangle\!=\!|-\!j\rangle$, the operator $\hat \varrho_f$ maps as}\\[-6mm]
\begin{equation}
\qquad\qquad \hat \varrho_f\ \mapsto \  \Pi\hat \varrho_f\Pi\!=\!\hat \varrho_g,
\end{equation}
\mbox{}\qquad  \qquad {\it where} \ $g(n,k)\!=\! f(-n,-k)$.\\
{\it Proof.} Since $\frak{g}_1(-n)\!=\!\frak{g}_1(n)$, under the transform $|j\rangle\mapsto \Pi|j\rangle\!=\!|-\!j\rangle$, 
\[
|n;k\rangle  \!=\!\sum\limits_{j=-s}^s|j\rangle\,
\mbox{\small $\frac{1}{\sqrt{\langle \frak{g}_1,\frak{g}_1\rangle}}$}\, {\rm e}^{-\frac{\pi {\rm i}}{d}nk}\,{\rm e}^{\frac{2\pi {\rm i}}{d}kj}\, \frak{g}_1(j\!-\!n),
\]
maps to
\[
\sum\limits_{j=-s}^s|\!-\!j\rangle\,
\mbox{\small $\frac{1}{\sqrt{\langle \frak{g}_1,\frak{g}_1\rangle}}$}\, {\rm e}^{-\frac{\pi {\rm i}}{d}nk}\,{\rm e}^{\frac{2\pi {\rm i}}{d}kj}\, \frak{g}_1(j\!-\!n)\!=\!|\!-\!n;-k\rangle .\qquad \Box
\] 

In the odd-dimensional case, the discrete Wigner function \cite{L,Vo,Wo} of a density operator \mbox{$\varrho \!:\!\!:\!\mathcal{H}\!\rightarrow\!\mathcal{H},$} is usually defined as $\mathfrak{W}_{\varrho}\!:\!\mathcal{R}\!\times\!\mathcal{R}\!\rightarrow\!\mathbb{R}$,
\begin{equation}
\mathfrak{W}_{\varrho}(n,k)\!=\!\frac{1}{d}\,\sum\limits_{m=-s}^s {\rm e}^{-\frac{4\pi {\rm i}}{d}km}\, \langle n\!+\!m|\varrho|n\!-\!m\rangle.
\end{equation}
 The discrete Wigner function of a pure state $\varrho\!=\!|\psi\rangle\langle \psi|$ is \cite{C1,C2,C3,CD1}
\begin{equation}
\mathfrak{W}_{\psi}(n,\!k)\!=\!\frac{1}{\mbox{\footnotesize $d$}}\!\sum\limits_{m=-s}^s \!{\rm e}^{-\frac{4\pi {\rm i}}{d}km}\, \psi(n\!+\!m)\, \overline{\psi(n\!-\!m)}.
\end{equation}
{\bf Theorem 11.}  {\it If the function $f\!:\!\mathcal{R}\!\times\!\mathcal{R}\!\rightarrow [0,d]$ is such that}
\begin{equation}
\sum\limits_{n,k=-s}^s f(n,k)\!=\!d,
\end{equation}
\qquad {\it then the discrete Wigner function of $\hat \varrho_f$ is}
\begin{equation}\fl
\qquad \qquad \mathfrak{W}_{\hat \varrho_f}(m,\ell )\!=\!C\sum\limits_{n,k=-s}^s f(n,k)\sum\limits_{\alpha,\beta=-\infty}^\infty (-1)^{\alpha \beta}{\rm e}^{-\frac{2\pi}{d}(m-n+\alpha \frac{d}{2})^2}{\rm e}^{-\frac{2\pi}{d}(\ell-k+\beta \frac{d}{2})^2},
\end{equation}
\qquad {\it where $C$  is a normalizing constant}.\\[3mm]
{\it Proof.} We have (see \cite{C2,C3,CD1})
\[\fl
\begin{array}{rl}
\mathfrak{W}_{\hat \varrho_f}(m,\ell )\!= & \!\!\!\mbox{\small $\frac{1}{d}$}\sum\limits_{n,k=-s}^s f(n,k)\, \mathfrak{W}_{|n;k\rangle}(m,\!\ell)\\
= & \!\!\!\mbox{\small $\frac{1}{d}$}\sum\limits_{n,k=-s}^s f(n,k)\, \mathfrak{W}_{|0;0\rangle}(m\!-\!n,\!\ell\!-\!k)\\
= & \!\!\!C\sum\limits_{n,k=-s}^s f(n,k)\sum\limits_{\alpha,\beta=-\infty}^\infty (-1)^{\alpha \beta}{\rm e}^{-\frac{2\pi}{d}(m-n+\alpha \frac{d}{2})^2}{\rm e}^{-\frac{2\pi}{d}(\ell-k+\beta \frac{d}{2})^2}.\qquad \Box
\end{array}
\]

\section{Composite quantum systems}
Let $s_A,s_B\!\in\!\{1,2,3,...\}$, \ $d_A\!=\!2s_A\!+\!1$, \   $d_B\!=\!2s_B\!+\!1$, \  $d\!=\!d_Ad_B$, \\
\mbox{}\quad \ \  $\mathcal{R}_A \!=\!\{-s_A, -s_A\!+\!1,...,s_A\!-\!1,s_A\}$, \  $\mathcal{H}_A\!=\!\mathbb{C}^{d_A}\equiv\{\, \psi\!:\!\mathcal{R}_A \!\rightarrow\!\mathbb{C}\, \}$, \\
\mbox{}\quad \ \ $\mathcal{R}_B \!=\!\{-s_B, -s_B\!+\!1,...,s_B\!-\!1,s_B\}$, \ $\mathcal{H}_B\!=\!\mathbb{C}^{d_B}\equiv\{\, \varphi\!:\!\mathcal{R}_B \!\rightarrow\!\mathbb{C}\, \}$, \\
\mbox{}\quad \ \ $\mathcal{R}\!=\!\mathcal{R}_A\!\times\!\mathcal{R}_B,$ \ $\mathcal{H}\!=\!\mathcal{H}_A\!\otimes\!\mathcal{H}_B\equiv\{\, \Psi\!:\!\mathcal{R}_A\!\times\!\mathcal{R}_B \!\rightarrow\!\mathbb{C}\, \}$. \\
The tensor product of two tight frames is a tight frame. \ Particularly, 
\[
 \left\{\, |n,\!m;\!k,\!\ell\rangle\!\equiv\!|n,\!m;\!k,\!\ell\rangle_{AB}\!=\!|n;\!k\rangle_{\!{}_A}\!\otimes\!|m;\!\ell\rangle_{\!{}_B}\ \left|\ \ 
n,k\!\in\!\mathcal{R}_A,\ \ m,\ell\!\in\!\mathcal{R}_B\,\right. \right\}
\]
is a tight frame in $\mathcal{H}\!=\!\mathcal{H}_A\!\otimes\!\mathcal{H}_B$, namely
\[\fl
\begin{array}{rl}
\mbox{\small $\frac{1}{d}$}\!\!\!\sum\limits_{n,k=-s_A}^{s_A}\sum\limits_{m,\ell=-s_B}^{s_B}  |n,\!m;\!k,\!\ell\rangle\langle n,\!m;\!k,\!\ell|\!= & \!\!\!\mbox{\small $\frac{1}{d}$}\!\!\!\sum\limits_{n,k=-s_A}^{s_A}\sum\limits_{m,\ell=-s_B}^{s_B}  |n;\!k\rangle_{\!{}_A}\!\otimes\!|m;\!\ell\rangle_{\!{}_B}\ {}_{{}_A}\!\langle n;\!k|\otimes\!{}_{{}_B}\!\langle m;\!\ell|\\
= & \!\!\!\mbox{\small $\frac{1}{d_Ad_B}$}\!\!\!\sum\limits_{n,k=-s_A}^{s_A}\sum\limits_{m,\ell=-s_B}^{s_B}  |n;\!k\rangle_{\!{}_A}\!\langle n;\!k|\!\otimes\!|m;\!\ell\rangle_{\!{}_B}\!\langle m;\!\ell|\\
= & \!\!\!\mbox{\small $\frac{1}{d_A}$}\!\!\!\sum\limits_{n,k=-s_A}^{s_A}|n;\!k\rangle_{\!{}_A}\!\langle n;\!k|\!\otimes\!\mbox{\small $\frac{1}{d_B}$}\!\!\!\sum\limits_{m,\ell=-s_B}^{s_B}  |m;\!\ell\rangle_{\!{}_B}\!\langle m;\!\ell|\\
= & \!\!\!\mathbb{I}_{\mathcal{H}_A}\!\otimes \mathbb{I}_{\mathcal{H}_B}\!=\!\mathbb{I}_{\mathcal{H}}.
\end{array}
\] 

By using the finite frame quantization, we associate the linear operator
\begin{equation}
\hat \varrho_f\!=\!\mbox{\small $\frac{1}{d}$}\!\!\!\sum\limits_{n,k=-s_A}^{s_A}\sum\limits_{m,\ell=-s_B}^{s_B} f(n,m;k,\ell)\, |n,\!m;\!k,\!\ell\rangle\langle n,\!m;\!k,\!\ell|
\end{equation}
to each function  
\begin{equation}
f\!:\!(\mathcal{R}_A\!\times\!\mathcal{R}_B)\!\times\!(\mathcal{R}_A\!\times\!\mathcal{R}_B)\!\rightarrow \![0,d],
\end{equation}
defined on the discrete phase space $\mathcal{R}^2$, and satisfying the relation
\begin{equation}
\sum\limits_{n,k=-s_A}^{s_A}\sum\limits_{m,\ell=-s_B}^{s_B} f(n,m;k,\ell)\!=\!d.
\end{equation}
If \ $f\!:\!\mathcal{R}_A\!\times\!\mathcal{R}_A\!\rightarrow [0,d_A]$  \ and \  $g\!:\!\mathcal{R}_B\!\times\!\mathcal{R}_B\!\rightarrow [0,d_B]$ \ are such that 
\begin{equation}
\qquad\qquad  \sum\limits_{n,k=-s_A}^{s_A} f(n,k)\!=\!d_A,\qquad 
\sum\limits_{m,\ell=-s_B}^{s_B} g(m,\ell)\!=\!d_B
\end{equation}
then
\[\fl
\quad \begin{array}{rl}
\quad \hat \varrho_f\!\otimes\!\hat \varrho_g\!= & \!\!\!\mbox{\small $\frac{1}{d_A}$}\sum\limits_{n,k=-s_A}^{s_A} f(n,k)\, |n;\!k\rangle_{\!{}_A}\!\langle n;\!k|\otimes \mbox{\small $\frac{1}{d_B}$}\sum\limits_{m,\ell=-s_B}^{s_B} g(m,\ell)\, |m;\!\ell\rangle_{\!{}_B}\!\langle m;\!\ell|\\
= & \!\!\!\mbox{\small $\frac{1}{d}$}\sum\limits_{n,k=-s_A}^{s_A}\sum\limits_{m,\ell=-s_B}^{s_B} f(n,k)\,g(m,\ell)\, |n;\!k\rangle_{\!{}_A}\!\langle n;\!k|\otimes   |m;\!\ell\rangle_{\!{}_B}\!\langle m;\!\ell|\\
= & \!\!\!\mbox{\small $\frac{1}{d}$}\sum\limits_{n,k=-s_A}^{s_A}\sum\limits_{m,\ell=-s_B}^{s_B} f(n,k)\,g(m,\ell)\, |n;\!k\rangle_{\!{}_A}\!\otimes\! |m;\!\ell\rangle_{\!{}_B}\, {}_{{}_A}\!\langle n;\!k|\!\otimes\! {}_{{}_B}\!\langle m;\!\ell|\\
= & \!\!\!\mbox{\small $\frac{1}{d}$}\sum\limits_{n,k=-s_A}^{s_A}\sum\limits_{m,\ell=-s_B}^{s_B} f(n,k)\,g(m,\ell)\, |n,\!m;\!k,\!\ell\rangle\langle n,\!m;\!k,\!\ell|\\
= & \!\!\!\hat \varrho_h,
\end{array}
\]
where \ $h\!:\!(\mathcal{R}_A\!\times\!\mathcal{R}_B)\!\times\!(\mathcal{R}_A\!\times\!\mathcal{R}_B)\!\rightarrow \![0,d], \  h(n,m;k,\ell)\!=\!f(n,k)\, g(m,\ell).$\\[3mm]
{\bf Theorem 12.} {\it If $f\!:\!(\mathcal{R}_A\!\times\!\mathcal{R}_B)\!\times\!(\mathcal{R}_A\!\times\!\mathcal{R}_B)\!\rightarrow \![0,d]$ is such that}
\begin{equation}
\qquad \quad \sum\limits_{n,k=-s_A}^{s_A}\sum\limits_{m,\ell=-s_B}^{s_B} f(n,m;k,\ell)\!=\!d,
\end{equation}
\mbox{}\\[-4mm]
\mbox{}\qquad \qquad \qquad  {\it then:}\\[-6mm]
\begin{equation}
\  a)\qquad {\rm tr}_{A}\, \hat \varrho_f\!=\!\hat \varrho_{f_B},
\end{equation}
\mbox{}\qquad \qquad \qquad {\it where}\quad 
$f_B\!:\!\mathcal{R}_B\!\times\!\mathcal{R}_B\!\rightarrow [0,d_B],\quad 
f_B(m,\!\ell)\!=\!\frac{1}{d_A}\sum\limits_{n,k=-s_A}^{s_A}\!\!\! f(n,\!m;\!k,\!\ell)$.
\begin{equation}
\ b)\qquad {\rm tr}_{B}\, \hat \varrho_f\!=\!\hat \varrho_{f_A},
\end{equation}
\mbox{}\qquad \qquad \qquad {\it where}\quad 
$f_A\!:\!\mathcal{R}_A\!\times\!\mathcal{R}_A\!\rightarrow [0,d_A],\quad 
f_A(n,\!k)\!=\!\frac{1}{d_B}\sum\limits_{m,\ell=-s_B}^{s_B}\!\!\! f(n,\!m;\!k,\!\ell)$.\\
{\it Proof.} a) We have
\[\fl
\quad \begin{array}{rl}
\quad {\rm tr}_{A}\, \hat \varrho_f\!= & \!\!\!\sum\limits_{a=-s_A}^{s_A}{}_{{}_A}\!\langle a|\hat \varrho_f|a\rangle_{\!{}_A} \\
= & \!\!\!\sum\limits_{a=-s_A}^{s_A}\mbox{\small $\frac{1}{d}$}\!\!\!\sum\limits_{n,k=-s_A}^{s_A}\sum\limits_{m,\ell=-s_B}^{s_B} f(n,m;k,\ell)\, {}_{{}_A}\!\langle a|n,\!m;\!k,\!\ell\rangle\langle n,\!m;\!k,\!\ell|a\rangle_{\!{}_A} \\
= & \mbox{\small $\frac{1}{d}$}\!\!\!\sum\limits_{n,k=-s_A}^{s_A}\sum\limits_{m,\ell=-s_B}^{s_B} f(n,m;k,\ell)\,\!\!\!\sum\limits_{a=-s_A}^{s_A} {}_{{}_A}\!\langle a|n;\!k\rangle_{\!{}_A}\!\langle n;\!k|a\rangle_{\!{}_A}\,|m;\!\ell\rangle_{\!{}_B}\!\langle m;\!\ell|\\
= & \mbox{\small $\frac{1}{d}$}\!\!\!\sum\limits_{m,\ell=-s_B}^{s_B}\sum\limits_{n,k=-s_A}^{s_A} f(n,m;k,\ell)\,\!\!\!\sum\limits_{a=-s_A}^{s_A} {}_{{}_A}\!\langle n;\!k|a\rangle_{\!{}_A}\!\langle a|n;\!k\rangle_{\!{}_A} \,|m;\!\ell\rangle_{\!{}_B}\!\langle m;\!\ell|\\
= & \mbox{\small $\frac{1}{d_B}$}\!\!\!\sum\limits_{m,\ell=-s_B}^{s_B}\mbox{\small $\frac{1}{d_A}$}\sum\limits_{n,k=-s_A}^{s_A} f(n,m;k,\ell)\,|m;\!\ell\rangle_{\!{}_B}\!\langle m;\!\ell|\\
= & \!\!\!\mbox{\small $\frac{1}{d_B}$}\sum\limits_{m,\ell=-s_B}^{s_B} f_B(m,\ell)\, |m;\!\ell\rangle_{\!{}_B}\!\langle m;\!\ell|.\\
\end{array}
\]
\  b)\quad \mbox{Similar to the proof of a)}.\qquad $\Box $\\

{\bf Theorem 13.} {\it In the case $d_A\!=\!d_B$, if $f\!:\!(\mathcal{R}_A\!\times\!\mathcal{R}_B)\!\times\!(\mathcal{R}_A\!\times\!\mathcal{R}_B)\!\rightarrow \![0,d]$ is such that}
\begin{equation}
\qquad \quad \sum\limits_{n,k=-s_A}^{s_A}\sum\limits_{m,\ell=-s_B}^{s_B} f(n,m;k,\ell)\!=\!d,
\end{equation}
\mbox{}\\[-4mm]
\mbox{}\qquad \qquad  {\it then, under the SWAP transform}
\begin{equation}
\mathcal{H}_A\!\otimes\!\mathcal{H}_B\!\rightarrow \!\mathcal{H}_A\!\otimes\!\mathcal{H}_B:\ |\varphi\rangle_{\!{}_A}\!\otimes\!|\psi\rangle_{\!{}_B}\!\mapsto \!|\psi\rangle_{\!{}_A}\!\otimes\!|\varphi\rangle_{\!{}_B},
 \end{equation}
\mbox{}\qquad \qquad  {\it the density operator $\hat \varrho_f$ maps as}\\[-6mm]
\begin{equation}
\qquad\qquad \hat \varrho_f\ \mapsto \  SWAP(\hat \varrho_f)\!=\!\hat \varrho_g,
\end{equation}
\mbox{}\qquad  \qquad {\it where} $g\!:\!(\mathcal{R}_A\!\times\!\mathcal{R}_B)\!\times\!(\mathcal{R}_A\!\times\!\mathcal{R}_B)\!\rightarrow \![0,d]$, \ $g(n,m;k,\ell)\!=\! f(m,n;\ell,k)$.\\
{\it Proof.} We have
\[\fl
SWAP(\hat \varrho_f)\!=\!\mbox{\small $\frac{1}{d}$}\!\!\!\sum\limits_{n,k=-s_A}^{s_A}\sum\limits_{m,\ell=-s_B}^{s_B} f(n,m;k,\ell)\, |m;\!\ell\rangle_{\!{}_A}\!\otimes\!|n;\!k\rangle_{\!{}_B}\, {}_{{}_A}\!\langle m;\ell|\!\otimes\!{}_{{}_B}\!\langle n;k|\!=\!\hat \varrho_g.\qquad \Box
\]
\section{Quantum channels obtained through finite frame quantization}
We continue to use the notations from the previous section and choose 
an auxiliary system $\mathcal{H}_{A'}$ such that ${\rm dim}\, \mathcal{H}_{A'}\!=\!{\rm dim}\, \mathcal{H}_{A}\!=\!2s_A\!+\!1$, and consequently $\mathcal{H}_{A'}\!=\!\{ \, \psi\!:\!\mathcal{R}_A\!\rightarrow\! \mathbb{C}\,\}$.
The pure quantum state 
\begin{equation}
\qquad\qquad |\Phi \rangle \!=\!|\Phi \rangle _{\!{}_{A'\!A}}\!=\!\mbox{\small $\frac{1}{\sqrt{d_A}}$}\sum\limits_{i=-s_A}^{s_A}  |i\rangle_{\!{}_{A'}}\!\otimes\!|i\rangle_{\!{}_A}\!=\!\mbox{\small $\frac{1}{\sqrt{d_A}}$}\sum\limits_{i=-s_A}^{s_A}  |ii\rangle
\end{equation}
is the most entangled state in $\mathcal{H}_{A'}\!\otimes\!\mathcal{H}_A$. In view of the channel-state duality (also called Choi-Jamiolkowski isomorphism), a quantum channel $\mathcal{E}\!:\!\mathcal{L}(\mathcal{H}_A)\!\rightarrow \!\mathcal{L}(\mathcal{H}_B)$ satisfying the relation 
$(\mathbb{I}\!\otimes \! \mathcal{E})(|\Phi\rangle\langle \Phi|)\!=\!\hat\varrho$
corresponds to each state $\hat \varrho \!:\!\mathcal{H}_{A'}\!\otimes\!\mathcal{H}_B\!\rightarrow \!\mathcal{H}_{A'}\!\otimes\!\mathcal{H}_B$, up to a normalization.
 Particularly, a quantum channel 
 $\mathcal{E}_f\!:\!\mathcal{L}(\mathcal{H}_A)\!\rightarrow \!\mathcal{L}(\mathcal{H}_B)$
corresponds to each state $\hat \varrho_f \!:\!\mathcal{H}_{A'}\!\otimes\!\mathcal{H}_B\!\rightarrow \!\mathcal{H}_{A'}\!\otimes\!\mathcal{H}_B$ with $f\!:\!(\mathcal{R}_A\!\times\!\mathcal{R}_B)\!\times\!(\mathcal{R}_A\!\times\!\mathcal{R}_B)\!\rightarrow \![0,d]$ satisfying
\begin{equation}
\qquad \quad \sum\limits_{n,k=-s_A}^{s_A}\sum\limits_{m,\ell=-s_B}^{s_B} f(n,m;k,\ell)\!=\!d.
\end{equation}
In the usual way, we prove that $\mathcal{E}_f$ admits the representation \cite{Pr}
\begin{equation}
\mathcal{E}_f(\hat \varrho)\!=\!\sum\limits_{n,k=-s_A}^{s_A}\sum\limits_{m,\ell=-s_B}^{s_B}K_{n,m;k,\ell}\ \hat \varrho \, K_{n,m;k,\ell}^\dag
\end{equation}
involving the Kraus operators $\hat K_{n,m;k,\ell}\!:\!\mathcal{H}_A\!\rightarrow \!\mathcal{H}_B$,
\begin{equation}
\hat K_{n,m;k,\ell}|i\rangle_{\!{}_A}\!=\!\sqrt{\mbox{\small $\frac{f(n,m;k,\ell)}{d}$}}\ {}_{{}_{A'}}\!\langle i|n,\!m;\!k,\!\ell\rangle.
\end{equation}
From the definition of $\hat K_{n,m;k,\ell}$ written in the form
\begin{equation}
\hat K_{n,m;k,\ell}|i\rangle_{\!{}_A}\!=\!\sqrt{\mbox{\small $\frac{f(n,m;k,\ell)}{d}$}}\sum\limits_{j=-s_B}^{s_B}|j\rangle_{\!{}_B}\langle ij|n,\!m;\!k,\!\ell\rangle
\end{equation}
we get the relation
\begin{equation}
 {}_{{}_B}\!\langle j|\hat K_{n,m;k,\ell}|i\rangle_{\!{}_A}\!=\!\sqrt{\mbox{\small $\frac{f(n,m;k,\ell)}{d}$}}\ \langle ij|n,\!m;\!k,\!\ell\rangle,
\end{equation}
whence
\begin{equation}
 {}_{{}_A}\!\langle i|\hat K_{n,m;k,\ell}^\dag|j\rangle_{\!{}_B}\!=\!\sqrt{\mbox{\small $\frac{f(n,m;k,\ell)}{d}$}}\ \langle n,\!m;\!k,\!\ell|ij\rangle
\end{equation}
and consequently
\begin{equation}
 {}_{{}_A}\!\langle i|\hat K_{n,m;k,\ell}^\dag\!=\!\sqrt{\mbox{\small $\frac{f(n,m;k,\ell)}{d}$}}\ \langle n,\!m;\!k,\!\ell|i\rangle_{\!\!{}_{A'}} .
\end{equation}
We have
\[\fl
\begin{array}{rl}
 (\mathbb{I}\!\otimes \! \mathcal{E}_f)(|\Phi\rangle\langle \Phi|)\!= & \!\!\!\mbox{\small $\frac{1}{d_A}$}\sum\limits_{i,j=-s_A}^{s_A}(\mathbb{I}\!\otimes \! \mathcal{E}_f)|ii\rangle\langle jj|\\
= & \!\!\!\mbox{\small $\frac{1}{d_A}$}\sum\limits_{i,j=-s_A}^{s_A}(\mathbb{I}\!\otimes \! \mathcal{E}_f)|i\rangle_{\!{}_{A'}}\!\langle j|\!\otimes\!|i\rangle_{\!{}_A}\!\langle j|\\
= & \!\!\!\mbox{\small $\frac{1}{d_A}$}\sum\limits_{i,j=-s_A}^{s_A}|i\rangle_{\!{}_{A'}}\!\langle j|\!\otimes\!\mathcal{E}_f(|i\rangle_{\!{}_A}\!\langle j|)\\
= & \!\!\!\mbox{\small $\frac{1}{d_A}$}\sum\limits_{i,j=-s_A}^{s_A}\sum\limits_{n,k=-s_A}^{s_A}\sum\limits_{m,\ell=-s_B}^{s_B}|i\rangle_{\!{}_{A'}}\!\langle j|\!\otimes\!\hat K_{n,m;k,\ell}|i\rangle_{\!{}_{A}}\!\langle j|\, \hat K_{n,m;k,\ell}^\dag\\
= & \!\!\!\mbox{\small $\frac{1}{d_A\, d}$}\!\sum\limits_{i,j=-s_A}^{s_A}\sum\limits_{n,k=-s_A}^{s_A}\sum\limits_{m,\ell=-s_B}^{s_B}\!f(n,m;k,\ell)\,|i\rangle_{\!{}_{A'}}\!\langle j|\!\otimes\! {}_{{}_{A'}}\!\langle i|n,\!m;\!k,\!\ell\rangle\langle n,\!m;\!k,\!\ell|j\rangle_{\!\!{}_{A'}} \\
= & \!\!\!\mbox{\small $\frac{1}{d_A}$}\sum\limits_{i,j=-s_A}^{s_A}|i\rangle_{\!{}_{A'}}\!\langle j|\!\otimes\! {}_{{}_{A'}}\!\langle i|\varrho_f|j\rangle_{\!{}_{A'}}\!=\!\mbox{\small $\frac{1}{d_A}$}\, \varrho_f \\
\end{array}
\]
because
\[
\begin{array}{rl}
 \langle nk|\sum\limits_{i,j=-s_A}^{s_A}|i\rangle_{\!{}_{A'}}\!\langle j|\!\otimes\! {}_{{}_{A'}}\!\langle i|\varrho_f|j\rangle_{\!{}_{A'}} |m\ell\rangle\!= & \!\!\!\sum\limits_{i,j=-s_A}^{s_A}\langle n|i\rangle\langle j|m\rangle\langle ik|\varrho_f|j\ell \rangle\\
= & \!\!\!\sum\limits_{i,j=-s_A}^{s_A}\delta_{ni}\,\delta_{jm}\, \langle ik|\varrho_f|j\ell \rangle\\
= & \!\!\!\langle nk|\varrho_f|m\ell \rangle .
\end{array}
\]
So, up to a normalization, we have $ (\mathbb{I}\!\otimes \! \mathcal{E}_f)(|\Phi\rangle\langle \Phi|)\!=\!\varrho_f$. In addition, 
\[\fl
\quad\begin{array}{l}
\sum\limits_{n,k=-s_A}^{s_A}\sum\limits_{m,\ell=-s_B}^{s_B}K_{n,m;k,\ell}^\dag K_{n,m;k,\ell}|i\rangle_{\!{}_A}\\
\quad \qquad =\!\sum\limits_{n,k=-s_A}^{s_A}\sum\limits_{m,\ell=-s_B}^{s_B}\sqrt{\mbox{\small $\frac{f(n,m;k,\ell)}{d}$}}\sum\limits_{b=-s_B}^{s_B}K_{n,m;k,\ell}^\dag|b\rangle_{\!{}_B}\langle ib|n,\!m;\!k,\!\ell\rangle\\
\quad \qquad =\!\sum\limits_{n,k=-s_A}^{s_A}\sum\limits_{m,\ell=-s_B}^{s_B}\sqrt{\mbox{\small $\frac{f(n,m;k,\ell)}{d}$}}\sum\limits_{b=-s_B}^{s_B}\sum\limits_{a=-s_A}^{s_A}|a\rangle_{\!{}_{A}}\!\langle a|K_{n,m;k,\ell}^\dag|b\rangle_{\!{}_B}\langle ib|n,\!m;\!k,\!\ell\rangle\\
\quad \qquad =\!\sum\limits_{n,k=-s_A}^{s_A}\sum\limits_{m,\ell=-s_B}^{s_B}\mbox{\small $\frac{f(n,m;k,\ell)}{d}$}\sum\limits_{b=-s_B}^{s_B}\sum\limits_{a=-s_A}^{s_A}|a\rangle_{\!{}_{A}}\, \langle n,\!m;\!k,\!\ell|ab\rangle\langle ib|n,\!m;\!k,\!\ell\rangle\\
\quad \qquad =\!\sum\limits_{n,k=-s_A}^{s_A}\sum\limits_{m,\ell=-s_B}^{s_B}\mbox{\small $\frac{f(n,m;k,\ell)}{d}$}\sum\limits_{b=-s_B}^{s_B}\sum\limits_{a=-s_A}^{s_A}|a\rangle_{\!{}_{A}}\, \langle ib|n,\!m;\!k,\!\ell\rangle\langle n,\!m;\!k,\!\ell|ab\rangle\\
\quad \qquad =\!\sum\limits_{b=-s_B}^{s_B}\sum\limits_{a=-s_A}^{s_A}|a\rangle_{\!{}_{A}}\, \langle ib|\hat \varrho_f|ab\rangle\\
\quad \qquad =\!\sum\limits_{b=-s_B}^{s_B}\sum\limits_{a=-s_A}^{s_A}|a\rangle_{\!{}_{A}}\, \langle ib|(\mathbb{I}\!\otimes \! \mathcal{E})(|\Phi\rangle\langle \Phi|)|ab\rangle\\
\quad \qquad =\!\sum\limits_{b=-s_B}^{s_B}\sum\limits_{a=-s_A}^{s_A}|a\rangle_{\!{}_{A}}\, \langle ib|\sum\limits_{j,\ell=-s_A}^{s_A}|j\rangle_{\!{}_{A'}}\!\langle \ell|\!\otimes\!\mathcal{E}_f(|j\rangle_{\!{}_A}\!\langle \ell|)|ab\rangle\\
\quad \qquad =\!\sum\limits_{b=-s_B}^{s_B}\sum\limits_{a=-s_A}^{s_A}|a\rangle_{\!{}_{A}}\, \sum\limits_{j,\ell=-s_A}^{s_A}\langle i|j\rangle\!\langle \ell|a\rangle\langle b|\mathcal{E}_f(|j\rangle_{\!{}_A}\!\langle \ell|)|b\rangle\\
\quad \qquad =\!\sum\limits_{a=-s_A}^{s_A}|a\rangle_{\!{}_{A}}\,{\rm tr}(\mathcal{E}_f(|i\rangle_{\!{}_A}\!\langle a|))\!=\!\sum\limits_{a=-s_A}^{s_A}|a\rangle_{\!{}_{A}}\,{\rm tr}(|i\rangle_{\!{}_A}\!\langle a|)\!=\!\sum\limits_{a=-s_A}^{s_A}|a\rangle_{\!{}_{A}}\,\delta_{ai}\!=\!|i\rangle_{\!{}_{A}},
\end{array}
\]
for any $i\!\in\!\mathcal{R}_A$, and consequently\\[-5mm]
\begin{equation}
\sum\limits_{n,k=-s_A}^{s_A}\sum\limits_{m,\ell=-s_B}^{s_B}K_{n,m;k,\ell}^\dag K_{n,m;k,\ell}\!=\!\mathbb{I}_{\!\mathcal{H}_A}.
\end{equation}

\section{Concluding remarks}
\mbox{The discrete coherent states (\ref{dcs}) approximate well \cite{CD2} the standard coherent states (\ref{ccs}).}
In the case of this finite frame, the use of the frame quantization seems to lead to a remarkable discrete version of certain linear operators \cite{CD2}.\\ 
Particularly, the density operators defined in this way have some significant properties, and may describe quantum states useful in certain applications.
%
\section*{References}

\end{document}